\documentclass[12pt, a4paper]{article}

\usepackage[english]{babel}
\usepackage[utf8x]{inputenc}
\usepackage{amsmath}
\usepackage{graphicx}
\usepackage[colorinlistoftodos]{todonotes}
\usepackage{soul}

\usepackage{rotating}
\usepackage{amsmath}
\usepackage{mathtools}
\usepackage{amsthm}
\usepackage{amsfonts}
\usepackage[dvips]{epsfig}
\usepackage{graphicx}
\usepackage{psfrag}

\usepackage{amssymb}
\usepackage{cancel}
\usepackage{pstricks} 
\usepackage{lscape}
\usepackage{cancel}
\usepackage{slashed}
\usepackage{pstricks} 
\usepackage{lscape}
\usepackage{color}
\usepackage{cite}
\usepackage{url}
\usepackage{caption}
\usepackage{bbm}
\usepackage{bm}
\usepackage{subfig}
\usepackage{bbold}
\usepackage{caption}

\newcommand{\beq}{\begin{equation}}
\newcommand{\eeq}{\end{equation}}
\newcommand{\ga}{\lower.7ex\hbox{$\;\stackrel{\textstyle>}{\sim}\;$}}
\newcommand{\la}{\lower.7ex\hbox{$\;\stackrel{\textstyle<}{\sim}\;$}}

\newcommand{\be}{\begin{equation}}
\newcommand{\ee}{\end{equation}}
\newcommand{\bea}{\begin{eqnarray}}
\newcommand{\eea}{\end{eqnarray}}

\makeatletter
\newcommand{\Cen}[2]{%
  \ifmeasuring@
    #2%
  \else
    \makebox[\ifcase\expandafter #1\maxcolumn@widths\fi]{$\displaystyle#2$}%
  \fi
}
\makeatother

\begin{document}

\begin{flushright}
{\tt KCL-PH-TH/2018-05}, {\tt CERN-PH-TH/2018-028}
\end{flushright}

\vspace{1cm}
\begin{center}
{\bf {\LARGE A Simple No-Scale Model of Modulus Fixing \\
\vspace {0.3cm}
and Inflation}}
\end{center}

\vspace{0.05in}

\begin{center}{
{\bf John~Ellis}$^{a,b}$,
{\bf Malcolm~Fairbairn}$^a$,
{\bf Antonio~Enea~Romano}$^{c}$ and
{\bf {\' O}scar~Zapata}$^{c}$
}
\end{center}

\begin{center}
{\em $^a$ Theoretical Particle Physics and Cosmology Group, Department of
  Physics, King's~College~London, London WC2R 2LS, United Kingdom}\\[0.2cm]
{\em $^b$ National Institute of Chemical Physics \& Biophysics, R{\" a}vala 10, 10143 Tallinn, Estonia; \\
Theoretical Physics Department, CERN, CH-1211 Geneva 23,
  Switzerland}\\[0.2cm]
{\em $^c$ Instituto de F{\' i}sica, Universidad de Antioquia, A.A.1226, Medell{\' i}n, Colombia}

\end{center}

\bigskip
\bigskip

\centerline{\bf {\large ABSTRACT}}
\vspace{0.5cm}
{We construct a no-scale model of inflation with a single modulus whose real and imaginary parts are fixed by {simple power-law} corrections to the no-scale K{\" a}hler potential. Assuming an uplift of the minimum of the effective potential, the model yields a suitable number of e-folds of expansion and values of the tilt in the scalar cosmological density perturbations and of the ratio of tensor and scalar perturbations that are compatible with measurements of the cosmic microwave background radiation.}

\vspace{0.5in}

\begin{flushleft}
February 2018
\end{flushleft}
\medskip
\noindent

\newpage

\section{Introduction}

Cosmological inflation \cite{guth,Starobinsky, MukhChib,LAS} provides one of the most promising arenas 
for probing physics close to the Planck scale, potentially even providing a window onto string theory. 
The effective energy scale during inflation may well be within a few orders of magnitude of the string scale, 
and in a wide class of inflationary models the excursion in the effective inflaton field is trans-Planckian. 
It is therefore natural to use string theory as an inspiration for the construction of such models, 
or at least to constrain the model-builders' imaginations~\cite{StringInflationModels}.

Consistent string models generally incorporate supersymmetry, and there are many practical reasons 
for supposing that supersymmetry may become apparent at some energy scale below that of inflation~\cite{LowScaleSUSY}. 
These considerations motivate the construction of supersymmetric models of inflation, which also offer 
advantages in rendering more natural the apparent hierarchy between the Planck scale and the energy scale during inflation \cite{Cries}. 
Since inflation is a cosmological scenario
that necessarily involves gravity, the most plausible supersymmetric framework for constructing models of inflation is actually supergravity \cite{nost}. 
Within this general framework, no-scale supergravity~\cite{no-scale,EKN,EKN2,LN} stands out~\cite{GL,KQ,EENOS,otherns,ENO13,no-scalereview}, 
since at the classical level it has a positive-semidefinite potential with flat directions that do not restrict field excursions \cite{no-scale}. 
Moreover, it emerges as the form of low-energy field theory derived from compactifications of string theory~\cite{Witten}.

The simplest no-scale supergravity model has a single complex field $T$ that
parametrizes a non-compact SU(1,1)/U(1) coset manifold with a 
K\"ahler potential $K = -3 \ln (T + T^*)$~\cite{no-scale,EKN}, and would correspond to the
volume modulus in a string compactification \cite{Witten}. It is a much-debated, very general and open, question 
how the values of the real and imaginary components of this and other compactification moduli {could be}
fixed dynamically in the low-energy physical vacuum~\cite{KKLT,EKNAvatars}~\footnote{{An alternative
would be to consider a scenario in which the quantum degree of freedom corresponding to $T - T^*$ is an
(almost) massless axion-like particle~\cite{Conlon}.}}. It is natural also to ask whether (some component) 
of the $T$ field could serve as the inflaton, and how this could be combined with whatever mechanism
that fixes dynamically the real and imaginary components of $T$.

In this paper we explore a possible common solution to these problems that {postulates} power-law modifications of the
leading-order K\"ahler potential of the form $\Delta K \; = \; {c_n}/{(T+T^*)^n} + {d_m}/{(T-T^*)^m}$,
{the first of which is rooted in our understanding of perturbative corrections to string compactifications}~\cite{DKL,RM}.
We show that, for suitable values of the powers $n, m$ and the correction parameters $c_n, d_m$,
there is a unique minimum of the effective potential $V < 0$ with fixed values of both the real and imaginary 
parts of $T$. Recognizing that the solution of the cosmological constant problem is unknown, we
assume that some unspecified uplifting mechanism raises the minimum of the effective potential
to $V \simeq 0$, and explore the possibility of successful inflation with the resulting positive
semidefinite potential $V(T)$. We find regions of initial conditions for the real and imaginary 
parts of $T$ that yield a number of e-folds $N_*$ and values of the scalar tilt parameter $n_s$ 
and the ratio of tensor to scalar perturbations $r$ that are highly compatible with the available data 
on the cosmic microwave background (CMB) data: $N_* \sim 55, n_s = 0.967$ and $r \sim 0.0007$ \cite{planck15}.
This model therefore provides a successful scenario for inflation in the context of a minimal
string-inspired no-scale supergravity model.

\section{The Effective Potential and Modulus Fixing}

We recall that an ${\cal N} = 1$ supergravity theory is specified~\cite{cremmer} by a Hermitian K\"ahler 
function $K$ and a holomorphic superpotential $W$ via the combination
\be
G \; \equiv \; K + \ln W + \ln W^* \, .
\label{sugra}
\ee
The K\"ahler function specifies the kinetic terms for the scalar fields:
\be 
K_i^{j^*} \equiv \frac{\partial^2 K}{\partial \phi^i \partial \phi^*_{j}} \, ,
\label{Kinetic}
\ee
where $K_i^{j^*} \equiv \partial^2 K/ \partial \phi^i \partial \phi^*_j$ is the K\"ahler metric, 
and the effective potential is
\begin{equation}
V \; = \; e^G \left[ \frac{\partial G}{\partial  \phi^i} K^i_{j^*}  \frac{\partial G}{\partial  \phi^*_j} - 3 \right] + \; {\rm possible} \; D-{\rm terms} \, ,
\label{effpot}
\end{equation}
and $K^i_{j^*}$ is the inverse of the K\"ahler metric.
In the following we study the simplest possible ${\cal N} = 1$ supergravity model with a single complex scalar field $T$,
and an exponential superpotential for $T$: $W(T) = e^{\lambda T}$.

\begin{figure}[h!]
\centering
 \includegraphics[width=12cm]{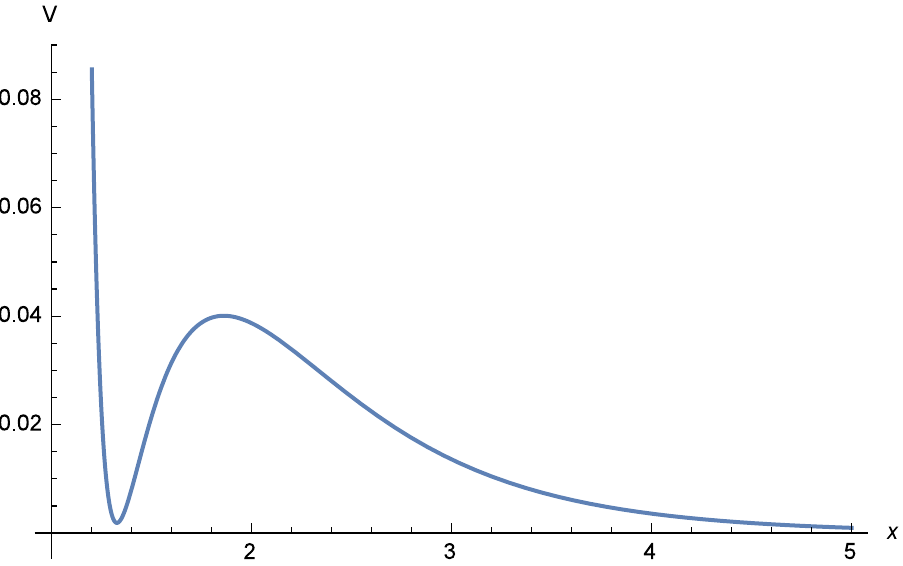}
 \caption{\it The effective potential (\protect\ref{effV}) as a function of $x = T + T^*$ for $n=4,\lambda=-1,c_n=3$.}
 \label{fig:Veff}
 \end{figure}

As mentioned in the Introduction, the minimal ${\cal N} = 1$ no-scale supergravity model has a K\"ahler potential $K = - 3 \ln (T + T^*)$~\cite{no-scale,EKN}.
We consider initially a K\"ahler potential with a correction of the form~\footnote{This form is inspired by the form of 
effective field theory found in~\cite{DKL} in describing (2, 2) vacua of the heterotic string.}:
\be
K \; = \; -3\ln\left(T+T^*\right)+\frac{c_n}{(T+T^*)^n} \, ,
\label{Vnoabs}
\ee
which yields a K\"ahler potential
\be
K_T^{T^*} \; = \; \frac{(T+T^*)^{n+2}}{c_n n (n+1)+3 (T+T^*)^n} \, . 
\label{realmodK}
\ee
In the following we denote the real part of $T$ by $x$ and the imaginary part by $y$: $T = x + i y$, and
define $g(x) \equiv K_T^{T^*}$. The resulting effective potential is
\be
V(x) \; = \;\frac{x^{-n-3} e^{c_n x^{-n}-\lambda  x} \left(c_n^2 n^2-c_n n x^n (3 n-2 \lambda  x-3)+\lambda  x^{2 n+1} (\lambda  x+6)\right)}{c_n n (n+1)+3 x^n} \, .
\label{effV}
\ee
The effective potential $V(x)$ has a local minimum at a non-zero value of $x$ when $n\geq 2$, 
as illustrated in Fig.~\ref{fig:Veff} for the specific choices $n=4,\lambda=-1,c_n=3$.

\begin{figure}[h!]
 \includegraphics[width=8cm]{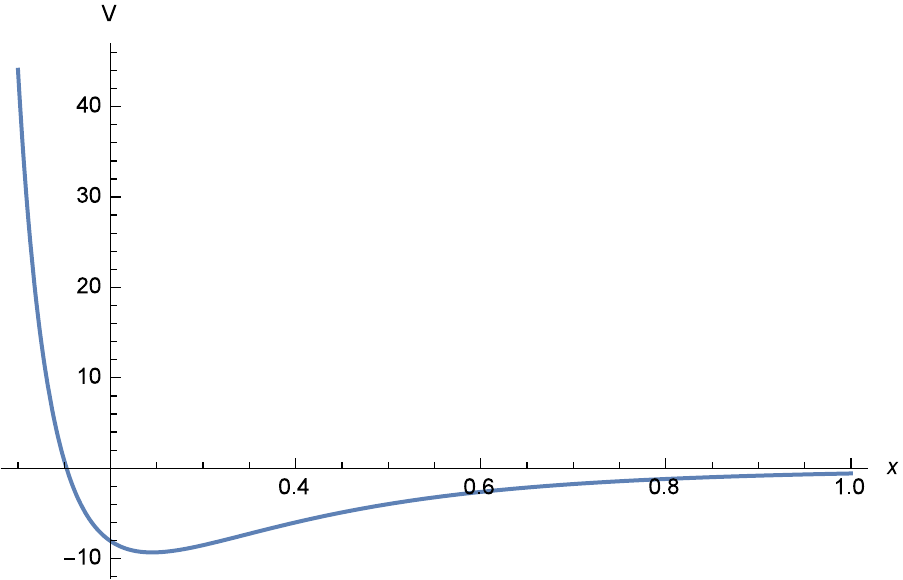}
 \includegraphics[width=8cm]{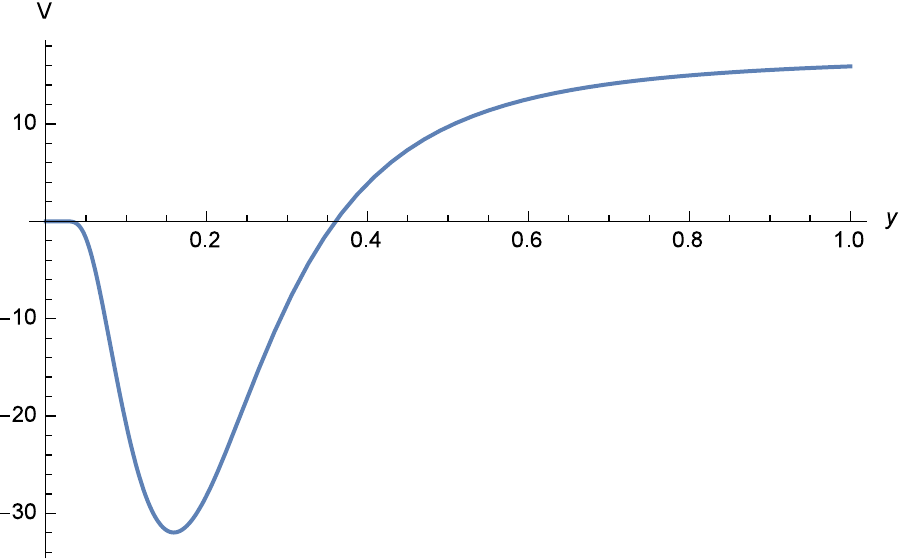}
 \caption{\it The effective potential obtained from (\ref{dmonly}) for $d_m=-0.05,m=3/2,\lambda=-1$ is plotted in the left panel
 as a function of $x$ for fixed $y=0.3$, and in the right panel as a function of $y$ for fixed $x=0.3$.}
 \label{fig:dmonly}
 \end{figure}

This is not the first example of stabilization of the real part of $T$ (see, for example, \cite{EKNAvatars}), but stabilization of the imaginary part
has proved more elusive (see, however, \cite{EGNO}).
In particular, the effective potential (\ref{effV}) is independent of $y$. In order to explore how $y$ may also be stabilized,
we {next consider adding instead} to the no-scale K\"ahler potential a dependence on the imaginary part of the modulus $T$
that is {also of power-law form, though not sharing its motivation from}
calculations of $\alpha^\prime$ corrections in string theory \cite{DKL,RM}:
\begin{equation}
K \; = \; -3\ln\left(T+T^*\right)+\frac{d_m}{|T-T^*|^m} \, .
\label{dmonly}
\end{equation}
In this case the effective potential takes the following form for $x, y > 0$:
\bea
V&=&\frac{1}{x^3}\exp{\Big[d_m y ^{-m}+\lambda  x\Big]}\Big[-3+ \nonumber \\
& & \hspace{-2cm} \frac{1}{\left(-d_m m (m+1) y ^{-m-2}+\frac{3}{x^2} \right)}\left(-d_m m \,  y ^{-m-1}+\lambda -\frac{3}{x}\right) \left(d_m m \,   y ^{-m-1}+\lambda \frac{3}{x}\right)\Big]  \,.
\label{Vxynocn}
\eea
Fig.~\ref{fig:dmonly} displays two slices through the effective potential (\ref{Vxynocn}) for $c_n=0,d_m=-0.05,m=3/2$ and $\lambda=-1$.
In the left panel we show an $x$ slice for fixed $y=0.3$, and in the right panel we show a $y$ slice for fixed $x=0.3$.
We see that in both slices there is a non-trivial minimum. We have also explored whether this example is suitable for inflation,
but found that this was not the case, and so do not consider further the $c_n = 0$ option.

\begin{figure}[h!]
\centering
\includegraphics[width=8cm,height=6cm]{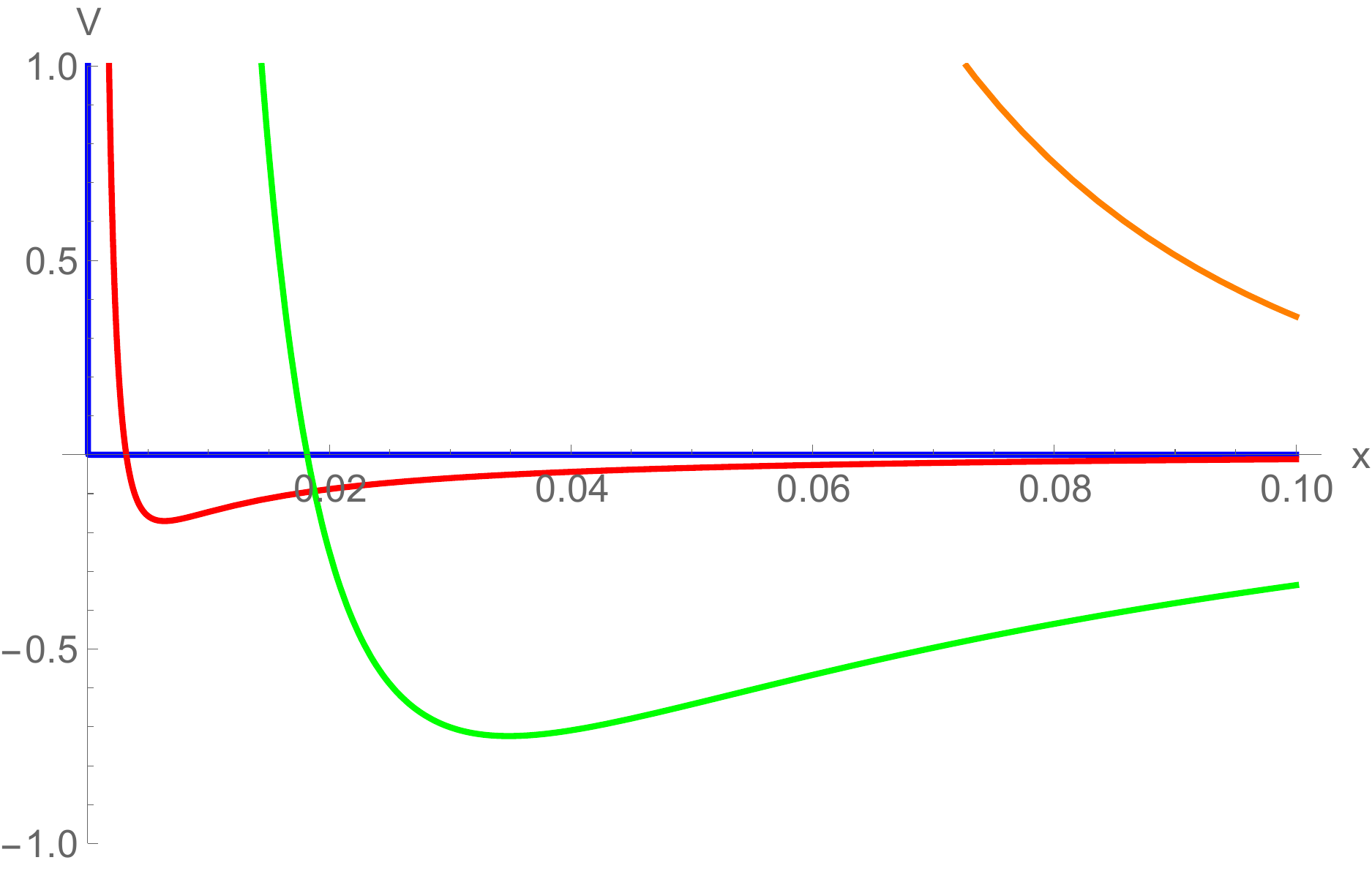}
\includegraphics[width=8cm,height=6cm]{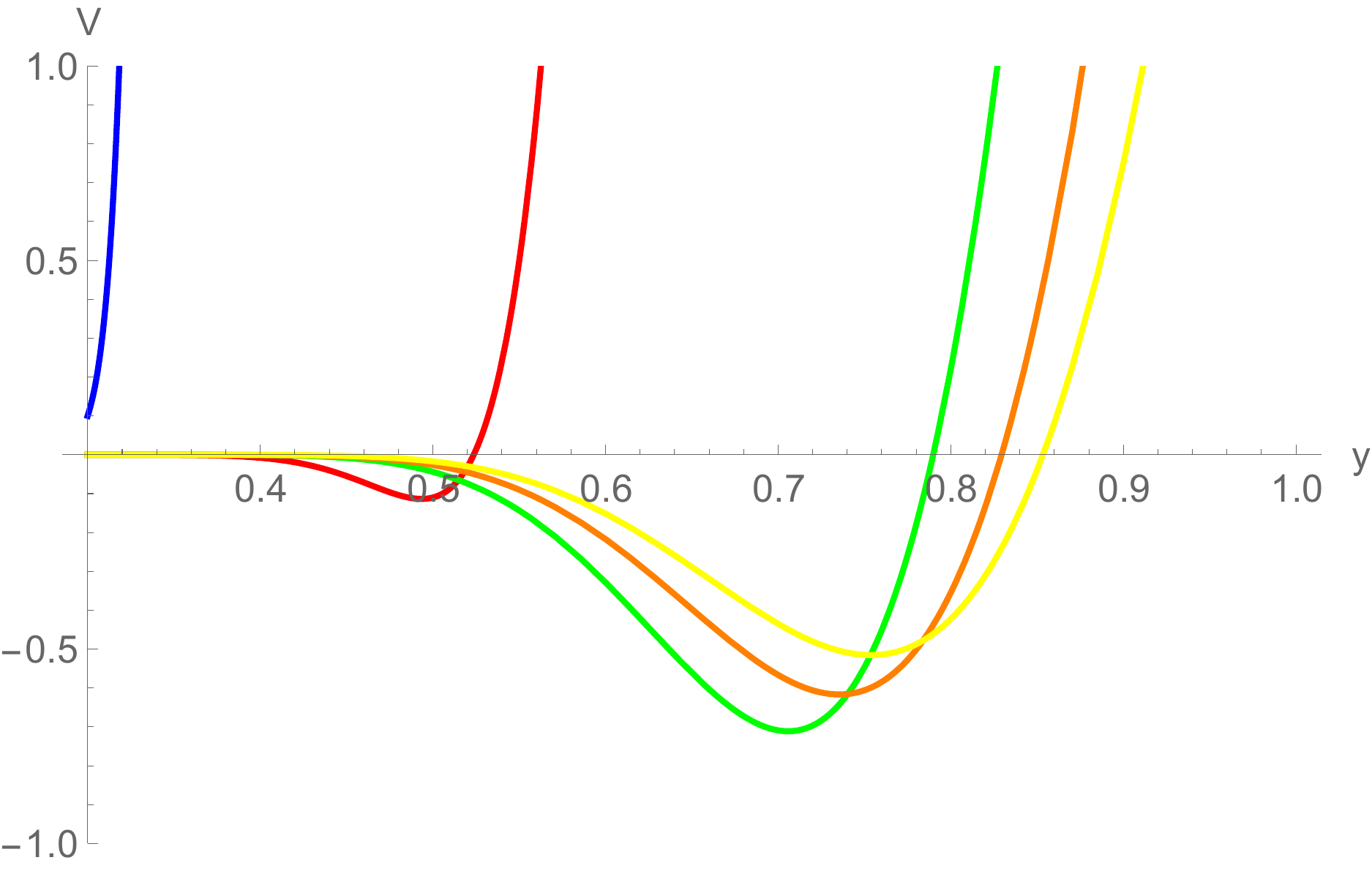}
\caption{\it The effective potential (\ref{Vxy}) for $n=-2, m=3/2, \lambda=-1,c_n = -5.9, d_m=-4.44$
is plotted in the upper panel as a function of  $x$, with the blue, red, green, and orange lines
corresponding to $y=\{0.3, 0.5, 0.7, 0.9\}$, and in the lower panel as a function of  $y$ with
the blue, red, green, orange and yellow lines corresponding to $x=\{6\, 10^{-6}, 0.004, 0.04, 0.06, 0.08, 0.1\}$.}
 \label{fig:slices}
\end{figure}

We have instead considered adding both the $T + T^*$-dependent term in (\ref{Vnoabs}) and the $T - T^*$-dependent term in (\ref{dmonly})
simultaneously to the no-scale K\"ahler potential:
\begin{equation}
K \; = \; -3\ln\left(T+T^*\right)+\frac{c_n}{|T+T^*|^n}+\frac{d_m}{|T-T^*|^m} \, .
\label{dm}
\end{equation}
In this case the effective potential takes the following form for $x, y > 0$:
\bea
V&=&\frac{1}{x^3}\exp{\Big[c_n x ^{-n}+d_m y ^{-m}+\lambda  x\Big]}\Big[-3+ \nonumber \\
& & \hspace{-2cm} \frac{1}{{c_n n (n+1) x ^{-n-2}-d_m m (m+1) y ^{-m-2}+\frac{3}{x^2}}}\left(-c_n n \,  x ^{-n-1}-d_m m \,  y ^{-m-1}+\lambda -\frac{3}{x}\right) \times \nonumber \\
&& \times \left(-c_n n \,   x ^{-n-1}+d_m m \,   y ^{-m-1}+\lambda \frac{3}{x}\right)\Big]  \,.
\label{Vxy}
\eea
Fig.~\ref{fig:slices} shows slices through the effective potential (\ref{Vxy})
for the choices $n=-2, m=3/2, \lambda=-1,c_n = -5.9, d_m=-4.44$.
The upper panel shows the $x$ dependence of the potential for several fixed values
of $y$ and the lower panel shows the $y$ dependence for several fixed values of $x$. We see that the real
component $x$ of the modulus $T$ is fixed at a non-zero value for the values $y=\{0.5, 0.7\}$, and that the 
imaginary component $y$ is fixed at a non-zero value for the values $x=\{0.04, 0.06, 0.08, 0.1\}$.

\begin{figure}[h!]
\centering
\includegraphics[width=10cm,height=10cm]{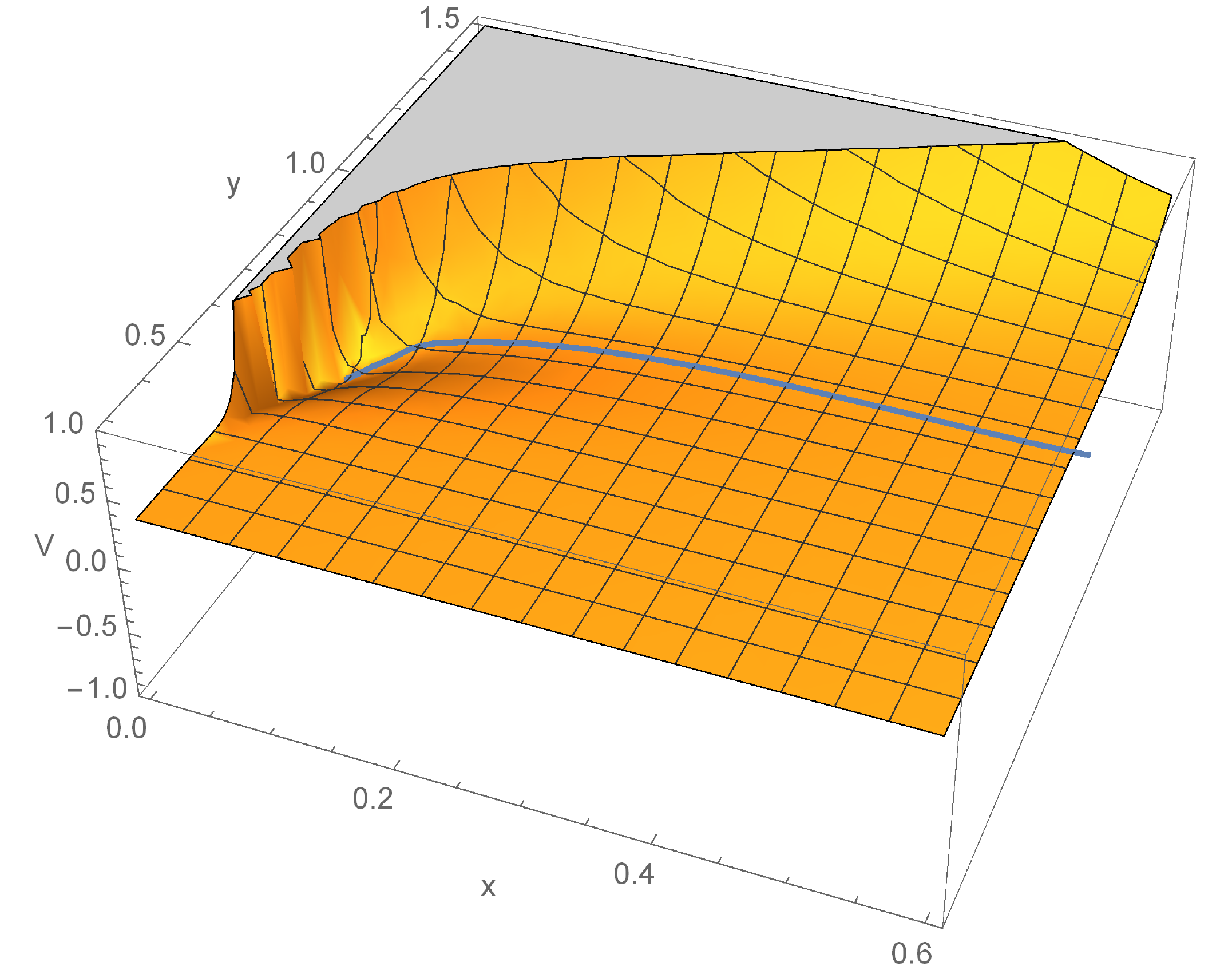}
\caption{\it The effective potential (\ref{Vxy}) is plotted as a function of $x$ and $y$ for $n=-2, m=3/2, \lambda=-1,c_n = -5.9, 
d_m=-4.44$. A global minimum with $x, y \ne 0$ is clearly present. Also shown as a blue line is a possible field trajectory.}
 \label{fig:VabsMin}
\end{figure}

Fig.~\ref{fig:VabsMin} shows a 3-dimensional image of the potential (\ref{Vxy}) for the same parameter choices
$n=-2, m=3/2, \lambda=-1,c_n = -5.9, d_m=-4.44$ used in Fig.~\ref{fig:slices}. This confirms that there is indeed
a global minimum of the potential with $x, y \ne 0$. Thus, the Ansatz (\ref{dm}) achieves the goal of fixing both
the components of the modulus field $T$.

\section{A Realization of Inflation}

The effective potential shown in Fig.(\ref{fig:VabsMin}) exhibits an extended flat region in addition to
the global minimum, and we now study whether there are field trajectories ending in the minimum
that are suitable for cosmological inflation. In order to check this, we need to solve the equations of 
motion for the modulus field components $x,y$ in an expanding Universe described by a Friedman-Robertson-Walker (FRW) metric
\be
ds^2=-dt^2+a(t)^2 d\overrightarrow{x}^2 \,,
\ee
corresponding to the action 
\be
S=\int\sqrt{-g}[ K_{TT^*} \partial_\mu T \partial^\mu T^*-V]d^4x \,.
\ee
in curved space. Assuming a homogeneous FRW background, only the time derivative survives
in the kinetic term and we obtain the following effective Lagrangian
\bea
L&=&a(t)^3[K_{TT^*}(x,y)(\dot{x}^2+\dot{y}^2)-V(x,y)] \,,
\eea
where
\bea
K_{TT^*}&=&  \left(c_n n (n+1) x^{-n-2}-d_m m (m+1) y^{-m-2}+e_p p (p+1) x^{-p-2}+\frac{3}{x^2}\right) \, . \nonumber
\eea
Combining with Einstein's equations for the scale factor $a(t)$, we get the following
system of differential equations that describe completely the field evolution:
\bea
\frac{d}{dt}\left(\frac{\partial L}{\partial\dot{x}}\right)-\frac{\partial L}{\partial x}&=&0 \,, \\
\frac{d}{dt}\left(\frac{\partial L}{\partial\dot{y}}\right)-\frac{\partial L}{\partial y}&=&0 \,, \\
H^2&=&\frac{1}{3}\Big[K_{TT^*}(x,y)(\dot{x}^2+\dot{y}^2)+V(x,y))\Big] \, .
\eea
A representative solution of these equations of motions is also shown in Fig.~\ref{fig:VabsMin}, as a blue line
that starts at $\{x, y \} = \{0.7, 1 \}$ and terminates at the global minimum.

\begin{figure}[h!]
\centering
\includegraphics[width=10cm,height=7.5cm]{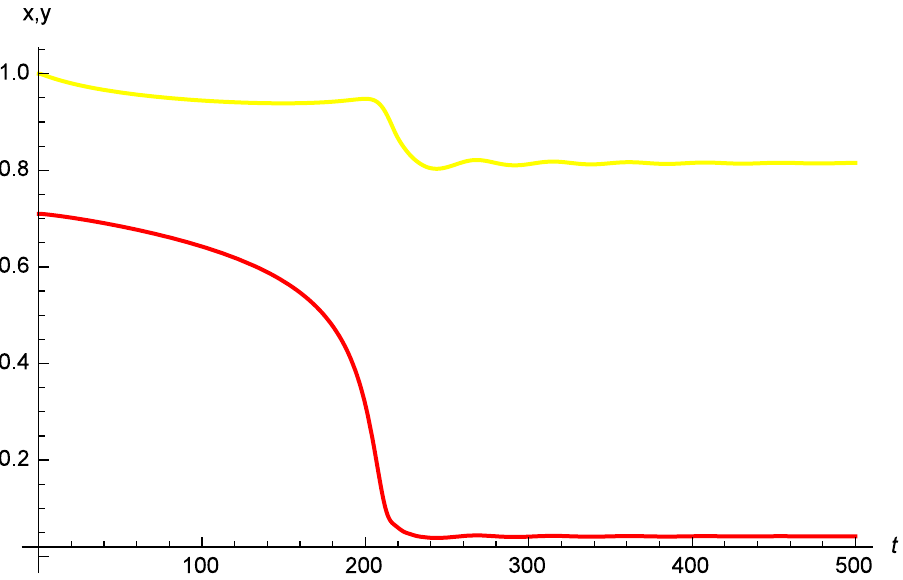}
\caption{\it The field components  $x$ and $y$ are plotted  in red and yellow, respectively,
as  functions of the time $t$ along the field trajectory shown in Fig.~\ref{fig:VabsMin}.}
 \label{fig:Yt}
\end{figure}

Fig.~\ref{fig:Yt} displays the evolutions along the field trajectory shown in Fig.~\ref{fig:VabsMin}
of the real and imaginary components $\{x, y \}$ of $T$ as functions of time, as red and yellow lines, respectively.
We see that $x$ decreases smoothly for $t \lesssim 240$, after which its value evolves only slowly, exhibiting
small oscillations. The value of $y$ changes by $\lesssim 10$\% for $t \lesssim 240$, after which it drops
to an almost constant value that also exhibits small oscillations.

\begin{figure}[h!]
\centering
\includegraphics[width=10cm,height=7.5cm]{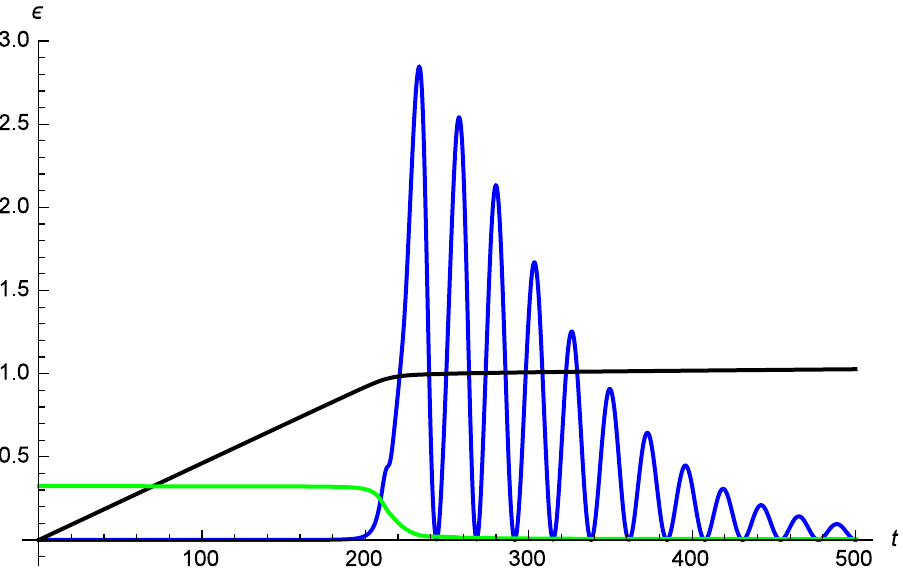}
\caption{\it The  Hubble parameter $H$, the slow-roll parameter $\epsilon$ and the  rescaled number of e-folds $N/70$ are plotted as  functions of time along the field trajectory shown in Fig.~\ref{fig:VabsMin} as green, blue and black lines, respectively.}
 \label{fig:Ht}
\end{figure}

Integrating the background equations we can compute the slow-roll parameters along the field trajectories, 
where we adopt the following definitions \cite{liddlelyth}:
\bea
\epsilon_1 &=& - \dot H/H^2 \,, \\
\epsilon_{i+1} &\equiv& \dot \epsilon_i/(H \epsilon_i) \,, \\
n_s - 1 &=& - 2 \epsilon_1 - \epsilon_2 - 2
\epsilon_1^2 \,, \\
r&=&16 \epsilon_1 \,.
\eea
Fig.~\ref{fig:Ht} displays the evolutions along the field trajectory shown in Fig.~\ref{fig:VabsMin}
of the Hubble parameter $H$ (green line), the 
slow-roll parameter $\epsilon_1$ (blue line), and the number of e-folds of expansion $N$ (black line, rescaled by
a reference value of 70). We see that the Hubble parameter varies only slowly until a time $t \simeq 200$,
falling to much smaller values when $t \gtrsim 240$. Correspondingly, the number of e-folds $N$ increases
nearly linearly until $t \sim 200$, after which it is nearly constant. The value of $\epsilon$ is small
until a similar value of $t$, after which it enters a period of damped oscillations with amplitudes that are
initially ${\cal O}(1)$. 

The initial conditions $x_i= 0.71,y_i=1, \dot{y_i}=\dot{x_i}=0$ lead to the field trajectory
shown in Fig.~\ref{fig:VabsMin}, which yields a number of e-folds $N \sim 55$, a scalar perturbation tilt
$n_s=0.967$ and tensor-to-scalar perturbation ratio $r=0.00069$, which are compatible with the
observational constraints~\cite{planck15}. Other choices of initial conditions also yield acceptable inflationary
trajectories. For example, changing the initial value of $y$ to $0.9$ but keeping the initial value of
$x$ fixed yields $n_s =0.961$ and $r= 0.00030$, whilst $y = 1.1$ yields $n_s =0.967$ and $r= 0.00075$,
also compatible with the observations. On the other hand, for initial values of $y < 0.7$
the non-triviality of the kinetic terms requires a deeper analysis than the approximate treatment that
is adequate for larger values of $y$.

\section{Conclusions}

We have presented in this paper a simple scenario for fixing both components of the modulus $T$ in the minimal no-scale
supergravity model with K\"ahler potential $K = -3 \ln (T + T^\dagger)$, which is {partly} based upon calculations of
theoretical calculations of corrections to this structure~\cite{DKL,RM}. In addition to yielding an effective potential that possesses
a well-defined, unique minimum, this simple model exhibits a plateau at larger values of the components of $T$.
We have found examples of field trajectories starting from initial values in this plateau region that yield
an acceptable number of e-folds of inflation and values of the CMB observables $n_s$ and $r$ that are 
compatible with observation~\cite{planck15}.

One interesting direction for future research will be to map out more completely the parameter space of
initial field values that are compatible with cosmological observations, {another will be to reconcile it
better with string considerations}, and another will be to
integrate this simple scenario into a framework with matter particles and
a scenario for reheating that would enable the number of inflationary e-folds to be calculated. In this
way, the simple model presented in this paper may serve as the kernel of a more complete cosmological
model.

\section*{Acknowledgements}

JE thanks Ruben Minasian for useful discussions.
{We thank Joe Conlon for pointing out en error in the first version of this paper.}
The work of JE and MF was supported partly by the STFC Grant ST/P000258/1. The work of JE was 
also supported partly by the Estonian Research Council via a Mobilitas Pluss grant and that of
MF was also supported by the European Research Council under the European Union's Horizon 2020 programme
(ERC Grant Agreement no.648680 DARKHORIZONS).
The work of AER was supported by the Dedicaci{\' o}n Exclusiva and Sostenibilidad programmes at the 
Universidad de Antioquia, as well as its CODI projects 2015-4044, 2016-10945 and 2016-13222, and the Colciencias mobility programme. 
The work of OZ was supported partly by Colciencias through the Grants 11156584269 and 111577657253.

  \end{document}